# Ultra-broadband Optical Diffraction Tomography

Martin Hörmann[1], Franco V. A. Camargo[2], Niek F. van Hulst[3,4], Giulio Cerullo[1,2], and Matz Liebel[3,5*]

[1]Dipartimento di Fisica, Politecnico di Milano, Piazza L. da Vinci 32, 20133 Milano, Italy

[2]Istituto di Fotonica e Nanotecnologie-CNR, Piazza L. da Vinci 32, 20133 Milano, Italy

[3]ICFO – Institut de Ciencies Fotoniques, The Barcelona Institute of Science and Technology, Av. Carl Friedrich Gauss, 3, 08860 Castelldefels, Barcelona, Spain

[4]ICREA – Institució Catalana de Recerca i Estudis Avançats, Passeig Lluís Companys 23, 08010 Barcelona

[5]Department of Physics and Astronomy, Vrije Universiteit Amsterdam, De Boelelaan 1081, Amsterdam, 1081 HV, The Netherlands

*email: m.liebel@vu.nl

**Abstract**

Optical diffraction tomography (ODT) is a powerful non-invasive 3D imaging technique, but its combination with broadband light sources is difficult. In this study, we introduce ultrabroadband ODT, covering over 150 nm of visible spectral bandwidth with a lateral spatial resolution of 150 nm. Our work addresses a critical experimental gap by enabling the measurement of broadband refractive index changes in 3D samples, a crucial information that is difficult to assess with existing methodologies. We present broadband, spectrally resolved ODT images of HeLa cells, obtained via pulse-shaping based Fourier transform spectroscopy. The spectral observations enabled by ultrabroadband ODT, combined with material-dependent refractive index responses, allow for precise three-dimensional identification of the nanoparticles within cellular structures. Our work represents a crucial step towards time and spectrally-resolved tomography of complex 3D structures with implications for life and materials science applications.

1. Introduction

Time-resolved spectroscopy studies the dynamics of light-induced processes, with applications ranging from fundamental photophysical processes over protein dynamics to complex devices and even single molecules[1–3]. It is a very active field of research as the ever-changing sample and material landscape calls for continuous innovation. Especially spatially resolved measurements have a tremendous impact on our understanding of, for example, nanoscale devices or biological materials. The combination of transient absorption spectroscopy with microscopy allows studying complex structure-function relationships with micro- to nanometre spatial and femtosecond temporal resolution, such as heat and carrier transport dynamics, uncovering exciting phenomena such as ballistic and hydrodynamic transport-regimes[4–6]. In parallel to the spectroscopic developments, spatially resolved pump-probe imaging, in the form of photothermal and phototransient approaches, is increasingly being used as a powerful imaging modality where pump-induced signals serve as a label-free means of contrast[7–13].

Most experiments provide spatially projected, and often ensemble-averaged, absorption changes (Figure 1a). However, samples are often three-dimensional and so is temporal information flow. In other words, existing techniques are unable to access the complex parameter space describing real samples, composed of spatial, temporal and spectral degrees of freedom. A technique that goes beyond the state-of-the-art would be highly desirable. Ideally, one would like to directly access the temporal evolution of the spectrally resolved complex refractive index (RI) and, further, connect this information with the underlying 3D nanostructure of the system of interest. Important steps towards realising this vision have been made[6,13–15] but a framework that enables transient complex RI spectroscopy with ultrabroadband/short pulses and nanometric spatial resolution on crowded 3D samples is still lacking. This shortcoming might, at first, be surprising as techniques such as optical diffraction tomography (ODT) readily provide 3D volumetric RI profiles, and even insight into the imaginary part of the RI, via computational synthesis based on multiple sample-projections[16–21]. Combined with an excitation pulse, transient ODT should therefore be in reach (Figure 1b).

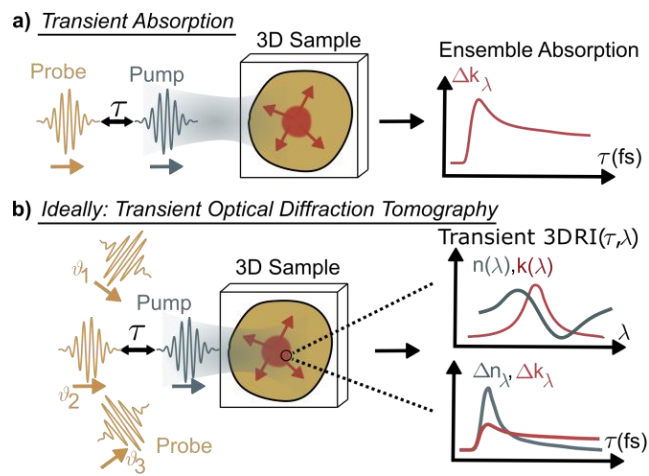

**Figure 1, Towards 3D transient absorption tomography.** a) Conventional transient absorption measures time-dependent absorption changes of a spatial ensemble-average. b) Optical diffraction tomography acquires an image stack at many different illumination, or probe, k-vectors to reveal the 3D complex RI distribution of a structured sample. A combination with transient excitation would enable the full characterization of the spatiotemporal response of any sample of interest.

A major hurdle towards realising transient measurements with ODT is the broad spectral bandwidth necessary to enable ultrafast observations. ODT requires knowledge of both the phase and the amplitude of all sample-projections, which is typically obtained holographically with spectrally narrow light. Non-holographic methods exist using intensity-only images[22–24] but those rely on computational post-processing that indirectly accesses the wave character of electric fields. Even though important steps towards increasing the bandwidth in holographic imaging have been taken, illumination bandwidths rarely exceed 10 nm[12–14,25–27], even for spectrally resolved ODT which requires wavelength sweeping[28–30]. This is a problem, since modern ultrafast transient absorption spectroscopy uses intrinsically ultra-broadband light pulses. Experimentally, it is challenging to combine ODT with ultrafast optics as the necessary broadband pulses exhibit extremely short temporal coherence lengths, far below 10 μm. Precise wavefront matching and ensuring correct temporal overlaps over large fields-of-observation and the entire wavelength range of a broadband pulse is highly non-trivial. The need to recover the spectral information of interest from images acquired in a colour-blind fashion further complicates the experiment.

Here, we implement and validate ultrabroadband ODT as a first step towards true 3D transient absorption microscopy or phototransient imaging. Using off-axis holography together with high numerical aperture (NA) based angle scanning we perform ODT using pulses with >150 nm bandwidths, supporting durations of approximately 5 fs. We further provide, and validate, a strategy that allows recovering spectral information based on Fourier transform spectroscopy. Our work bridges the gap between three-dimensional imaging, via ODT, with ultrabroadband pulses exhibiting transform limited durations in the <10 fs range. Integrating our strategy with pre-compressed pump and probe pulses, using existing solutions[12,31], has the potential to directly access ultrafast 3D observations .

The paper is organised as follows: we first give a detailed description of the experimental setup, addressing pitfalls, problems and solutions when using broadband pulses for holographic imaging. We then discuss the general data-processing workflow, from raw holograms to spectrally resolved 3D RI representation. Following these experimental aspects, we validate that spectral interference over the entire bandwidth is achieved. We then present proof-of-principle spectrally resolved tomograms of HeLa cells. Finally, we conclude by performing two-colour, spectrally multiplexed, ODT for nanoparticle identification in complex biological samples.

## 2. Methods

Spectrally resolved, ultrabroadband, ODT requires: i) a dedicated experimental setup suitable for temporally ultrashort light; ii) a means of recovering spectral information; iii) computational processing to recover the spectrally resolved 3D information. In the following, we will provide detailed information addressing these three key aspects.

### 2.1. Optical diffraction tomography via broadband off-axis holography

Figure 2a depicts the ultrabroadband ODT imaging system based on the 480-670 nm output of a supercontinuum laser (*SuperK EXTREME*, *NKT Photonics*), mimicking the spectral range of typical Yb-laser pumped broadband noncollinear optical parametric amplifiers[32,33]. The light enters a Mach-Zehnder interferometer that we, conceptually, divided into a signal and a reference arm to facilitate the discussion. Finally, a camera (*Q-2HFW, Adimec*) records the interference between the fields passing through the respective interferometric arms.

In more detail, the signal path contains a transmission microscope composed of a NA=1.2 (*Olympus UP-LSAPO60XW*, 60X, water immersion) and a NA=1.25 (*Olympus UPLSAPO40XS*, 40X, silicone immersion) objective for illumination and collection, respectively. An imaging system composed of three lenses with effective focal lengths of 100, 100 and 240 mm (visible achromats, *Thorlabs*) and the illumination objective conjugate a rotating grating (Ronchi type, 20 grooves/mm, *Edmund Optics*) with the sample plane. A hard-aperture, placed in a Fourier plane of the grating, blocks all but the first diffraction order and allows interrogating the sample at precisely defined, and adjustable, wave-vectors. The second microscope objective collects all the transmitted light and a 500 mm lens (visible achromat, *Thorlabs*) forms an image of the sample on the camera.

The reference arm contains bulk material plus a pair of 2° fused silica wedges (*LAYERTEC GmbH*) for precise chirp management. A translation stage matches the path length difference between the two interferometer arms. A 10x telescope expands the beam and eliminates minor wavefront curvature mismatch between signal and reference. A grating (19.5 grooves/mm), conjugated with the camera via a 1:1 imaging system, generates the reference wave required for off-axis holography, as its first diffraction order. Adequate beam-blocks placed into the Fourier plane of the grating eliminate all other diffraction orders.

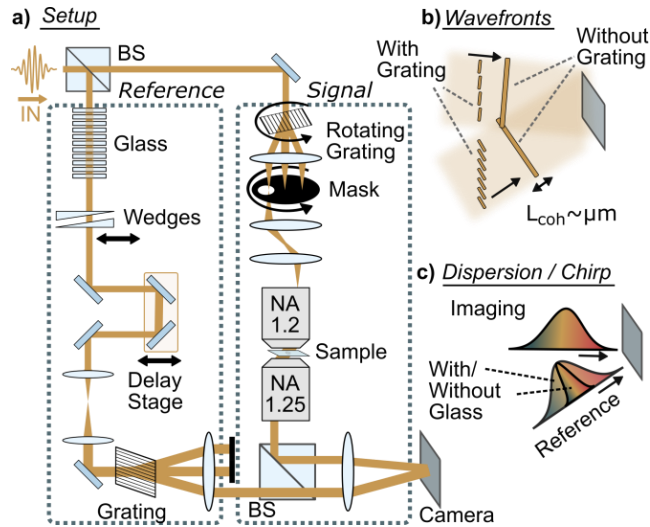

**Figure 2, Broadband ODT setup and experimental considerations.** A) Supercontinuum-based broadband high-NA ODT setup relying on angle-scanning via grating rotation; BS: beam-splitter. B) Gratings in the sample and reference arms ensure parallel wavefronts and hence optimum interference. C) Dispersion matching is necessary to ensure broadband interference.

The setup outlined above ensures broadband off-axis interference between the signal and refence arms by meeting two key requirements. Firstly, wavefront matching is ensured by performing angle scanning and off-axis holography using a diffraction grating which allows interfering temporally short waves on large 2D detectors[34,35] (Figure 2b). Secondly, interference over the entire spectral bandwidth is enabled by carefully matching the spectral phases, or chirps, of signal and reference beams at the camera (Figure 2c), which is possible via spectral interferometry[36] (Supplementary Information 1). Importantly, in this proof-of-concept work we control the relative spectral phase difference between the two fields and not the absolute spectral phase of the fields, which would be required for ultrafast 3D transient absorption microscopy. This reduces the experimental complexity as compared to operating with transform-limited pulses.

In a typical experiment, we record 60 off-axis holograms while systematically changing the illumination k-vector, via grating rotation, in steps of 6°. The laser repetition rate was 15.59 MHz and the camera integration time 1 ms to ensure that potential low-frequency vibrations do not degrade the interference contrast. In all experiments, we recorded sample and background information by capturing image stacks of objects of interest alongside identical stacks of empty regions, respectively.

## 2.2. Spectral Resolution

The imaging system described in Figure 2 enables broadband ODT but lacks spectral resolution. We provide this much-needed capability via pulse-shaping to perform Fourier-transform spectroscopy[1], using a simple Fourier filter, in combination with a spatial light modulator (SLM), placed between the supercontinuum laser and the ODT setup (Figure 3a). Our implementation is based on the established mirrored 4f-design in which a grating disperses the incoming light which is then Fourier transformed by a lens onto the liquid-crystal mask of the SLM (*Jenoptik SLM- S640d*) followed by back-reflection through the system. A polarizer placed in front of the SLM allows phase and amplitude control in a straight-forward manner.

Time-domain Fourier transform measurements[1] via a so-called "pulse pair" (Figure 3b) are a sampling strategy that is especially popular in ultrafast spectroscopy as it favourably integrates with the available broadband light sources[37]. In brief, the SLM modulates the frequencies, $\nu$, of the dispersed spectrum according as $M(\nu, \tau) = |\cos((\nu - \nu_0) * \pi * \tau)|^2$, with $\nu_0$ being the carrier frequency that generates a pulse pair with effective time delay $\tau$. Operating in a rotating frame[38,39] allows Nyquist-sampling the $\tau$ = 0-40 fs temporal range in 24 steps. A pixelwise one-dimensional fast Fourier transform (FFT) along the SLM-generated time-delay axis retrieves the spectral components of interest.

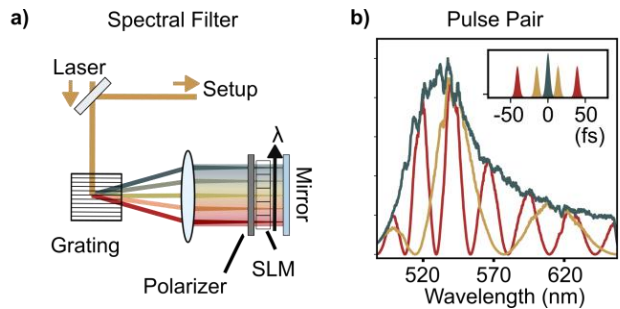

**Figure 3, Spectral imaging configuration.** a) A 4f-grating zero-dispersion pulse shaper equipped with a polarizer and spatial light modulator (SLM) in the back-focal plane manipulates the pulse spectrum. b) Selected spectra used for time-domain Fourier transform, or pulse-pair, imaging at a nominal spectral resolution of 6.25 THz, or 6.9 nm, at 575 nm. The inset shows a schematic sketch of the intensity of the DC pulse together with the time-delayed satellite pulse pairs.

## 2.3. Reconstruction Procedure

The setup outlined above yields datasets containing spectrally resolved sample-holograms, acquired at each illumination angle, alongside a second background dataset, acquired in an empty region. These data allow reconstructing the spectrally resolved 3D RI of the sample. Out of the many different reconstruction strategies[19,40–43] available, we chose the Rytov approximation[16,17,20,44], a widely used approach that enables straight-forward 3D reconstructions of the real part of the RI based on a set of complex 2D recordings. The reconstruction workflow, from raw holograms to 3D volumetric images, is schematically outlined in Figure 4 and explained in detail in the Supplementary Information 2. Before performing the actual reconstruction, we extract the complex interference terms from the image stacks, following established Fourier-filtering based[45] hologram-processing routines (Figure 4a). We retrieve the complex field, defined as $f_0(x, y, \tau, \vartheta)$, with $x$ and $y$ being the spatial coordinates, $\vartheta$ the illumination angle and $\tau$ the delay between the satellite pulses. We apply a one-dimensional pixelwise FFT to extract the wavelength dependent field from the temporal interferogram to yield complex fields $f_0(x, y, \lambda, \vartheta)$ as a function of wavelength and angle (Supplementary Information 3). The same procedure is applied to the background holograms yielding $f_{bg}(x, y, \lambda, \vartheta)$ which allows eliminating amplitude and phase contributions that do not originate from the sample, such as non-uniform illumination profiles or residual, static, phase differences between signal and reference waves. The background fields are, further, used to retrieve the illumination k-vectors, $\underline{k}_0 = (k_{x0}, k_{y0}, k_{z0})$, for each angle and wavelength (Supplementary Information 4).

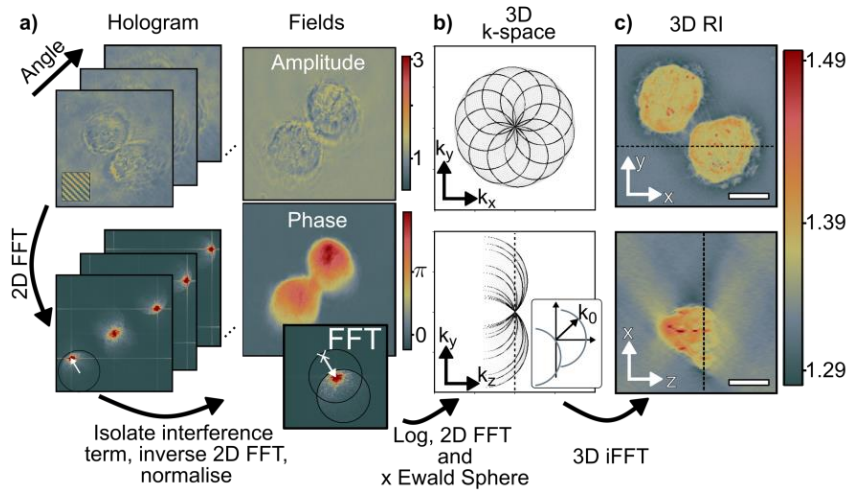

**Figure 4, From raw holograms to 3D tomograms.** a) A set of holograms (with and without sample) is recorded at different illumination angles and then Fourier-processed to retrieve normalized amplitude and phase images. The top inset shows the interference fringes of the hologram. The bottom inset corresponds to the 2D FFT of the normalized image. The white arrows show the (lateral) illumination beam in k-space and the shift due to normalization. b) The object's scattering potential in 3D k-space is sampled using the 2D FFT of the normalized fields and the corresponding Ewald sphere of each illumination angle. Only a few subsets are highlighted for clarity. The inset shows the shift of the Ewald sphere for an illumination with k-vector $\underline{k}_0$. c) An inverse 3D FFT of the Ewald sphere retrieves the 3D, real space, RI of the object. Dotted lines represent the respective image slices. Scale bar: 10 µm.

Following amplitude and phase image retrieval we move on to reconstructing the 3D volumes, processing each wavelength separately, following refs [17,20]. In simple terms, illuminating the sample at different angles yields projections that encode 3D information into a complex 2D image stack. Knowledge of the illumination angle allows placing this 2D information into a, computationally generated, 3D k-space on a so-called Ewald sphere. Once the entire information, for all angles but at a fixed wavelength, has been inserted, an inverse FFT yields the 3D real space image, or RI, for one wavelength component. Figure 4 summarises the entire workflow from raw holograms of fixed HeLa cells to xy- and xz-cuts of the tomogram. Supplementary Information 5 provides a short discussion of the artefacts in ODT due to the missing cone problem and of the choice of the Rytov approximation.

## 3. Results and applications

Our method retrieves spectrally resolved tomograms via Fourier transform spectroscopy but might suffer from potential experimental artefacts due to, for example, loss of interference as a result of residual spectral phase or spatial misalignment. To validate the approach we compared the electric fields, as retrieved from the broadband Fourier-transform holographic measurements, to an established, narrowband, slit-scan approach (Supplementary Information 6)[28–30]. The slit scan reconstructs broadband images from many narrowband observations at varying wavelengths. As such, it conveniently eliminates potential sources of artefacts but is, from a temporal-resolution perspective, incompatible with ultrafast observations. Irrespective, the two modalities deliver comparable spectral observations even though the temporal coherence lengths, $l_{coh} = \frac{c}{\Delta \nu}$, of individual observations differ by more than an order of magnitude: an ideal handle to ensure artefact-free observations.

Figure 5a shows raw intensity images, obtained for one illumination angle, as extracted from holograms recorded on HeLa cells. The unnormalized intensity images directly report on the intensity of the interference term, $A_s A_r$, with $A_s$ and $A_r$ being the signal and reference amplitudes, respectively. The images show a strong wavelength dependence, which should, in principle, reflect the spectrum of the supercontinuum source in the case of perfect interference over the entire spectrum.

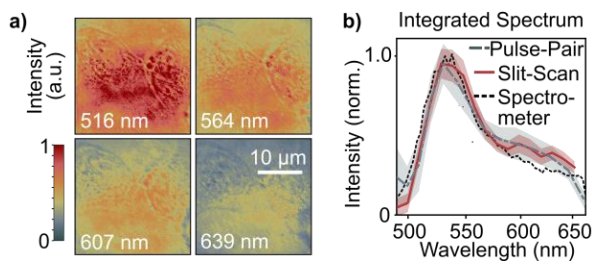

**Figure 5, Proof of broadband interference.** a) Intensity ($A_s A_r$) images underlying the HeLa cell sample reported in Figure 6 shown for one illumination angle at representative wavelengths obtained via pulse-pair imaging. b) Normalized integrated intensity images over 60 angles obtained via pulse pair imaging (dashed line), via a sequence of narrowband images (red line) and normalized source spectrum as measured by a spectrometer (dotted line). The shaded areas correspond to two standard deviations.

To benchmark the quality of our data, we spectrally integrate the intensity images and then combine all illumination angles into one data point, for each wavelength for both the Fourier transform approach and the slit scan. Further, we acquire ground-truth spectra using a commercial spectrometer (*Ocean Optics*). Figure 5b compares the normalized spectral intensities obtained using the three approaches. Overall, we observe near-perfect agreement between the slit scan and pulse pair methods. Minor discrepancies with respect to the spectrometer ground-truth data towards the red edge of the spectrum are apparent, which we attribute to potentially non-perfect quantum- or grating-efficiency calibrations of either the imaging camera or the spectrometer as well as marginal differences between beamsplitter reflectivity and transmissivity. Overall, the data suggest artefact-free ultra-broadband tomographic observations.

An important application of ODT is imaging biological samples, such as mammalian cells. To this end, Figure 6 presents spectrally resolved broadband tomography experiments, performed on HeLa cells and acquired with the pulse pair approach. Figure 6a shows xy- and yz-tomogram projections obtained for two fixed HeLa cells at three Fourier-extracted wavelengths of 516 nm, 564 nm and 607 nm. We observe the typical morphology for adherent cells alongside the expected RI differences between nucleus and cytoplasm. The small regions of considerably higher RI might be endosomes, lipid droplets or otherwise aggregated biological material. The seemingly void regions, with RI-values comparable to water, are most likely due to fixation-induced disruption of the structural integrity of the cells, a problem of paraformaldehyde-fixation that is especially visible with holographic and tomographic observations[46]. Overall, pulse-pair based ODT yields broadband, spectrally resolved, tomograms whose quality is comparable to those acquired via state-of-the-art narrowband approaches[47]. Spectrally, we only observe minor differences between the images, with a slight noise-increase towards the red part of the spectrum. Figure 6b confirms this notion by comparing RI measurements for the three wavelengths along representative x- and z-cuts. These observations are consistent with the fact that the RI of biological materials changes slowly in the visible spectral range. Further, it can be observed that the RI drops below the RI of water at the edges of the cells, which is due to the well-known missing cone artefact (Supplementary Information 5). The overall agreement of the RI between all wavelengths allows generating wavelength-averaged tomograms as the RI mean. Figure 6c shows an averaged RI-distribution obtained for nine distinct wavelengths in the 516 nm – 607 nm observation window. Importantly, such averaging is only possible once the 3D real space RI distributions are obtained, due to the wavelength dependent nature of the k-vectors.

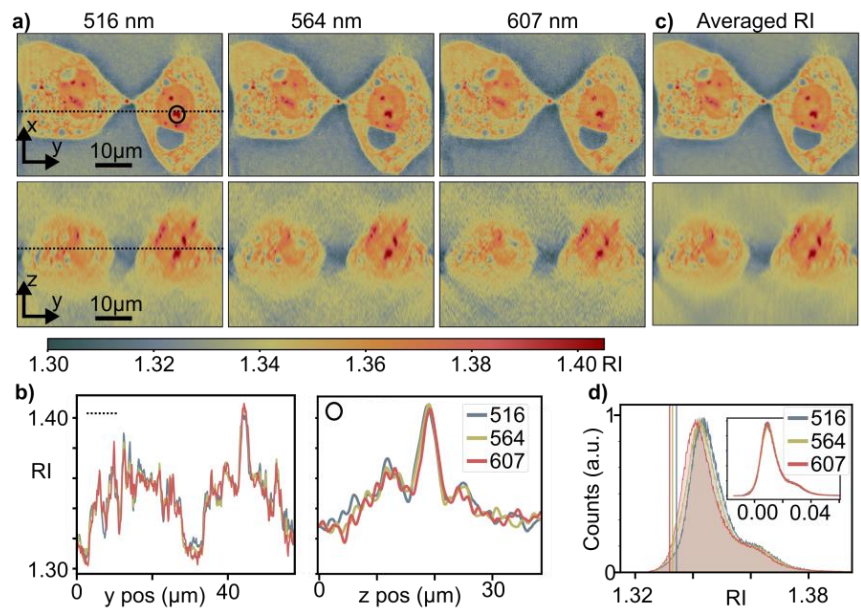

**Figure 6, Wavelength resolved 3D tomography of HeLa cells.** a) xy- and xz-projections of Hela cells imaged with the pulse-pair method and reconstructed for different wavelengths. The lines correspond to the respective cut in the xy- and yz-images. b) RI along y and z (solid line and red circle in a), respectively). c) The averaged RI obtained using nine reconstructions in the wavelength range of 516 nm to 607 nm. d) RI distribution for the HeLa cell located on the right side of the images shown in a), the vertical lines indicate the RI of the surrounding medium (water). The inset shows the histograms with the corresponding surrounding medium subtracted.

To gain more quantitative insight, we compare the RI distributions of the entire right HeLa cell for different wavelengths with that of water (Figure 6d). We observe RI changes as a function of wavelength with a slight decrease for increasing wavelengths. Importantly, the histograms show shifts similar to water, as expected given that off-resonant organic matter and water exhibit comparable chromatic dispersion in our spectral range. This observation suggests that the Rytov approximation, albeit underestimating the RI values, is spectrally accurate. The minor differences in the histograms, after normalizing them to their respective water RI (inset Figure 6d), is most likely due to the wavelength dependent spatial resolution which slightly reduces the apparent RI for small objects contained in the lower RI surrounding water (Supplementary Information 7).

Figure 6 highlights the ability of ultra-broadband ODT to deliver high-quality, spectrally accurate, 3D reconstructions. To take first steps towards interrogating spectrally resonant systems, we incubate HeLa cells with Au nanoparticles (NPs). The representative 3D response of an isolated single 150 nm AuNPs at 505 nm and 580 nm (Figure 7a) suggests that spectrally resolved ODT should be well-suited for identifying such particles in scattering biological matter. To test this hypothesis, we employ a tomographic implementation of a multiplexed holographic detection scheme, based on our previous design[48]. Our approach simultaneously records tomograms at 505 nm and 580 nm in order to identify AuNPs within cells. Figure 7b summarizes the necessary experimental modifications to implement two color-detection. Prior to the imaging system we select the two colors, 505 nm and 580 nm, from the broadband light source using a hard-aperture Fourier filter in front of the SLM (Figure 3). Multiplexed two-color interference is ensured by selectively blocking the respective colors after the first lens following the 2D grating in the reference path. Blocking one of them in each reference beam, respectively, retrieves two references with different colors and k-vectors. The (complex) images of the two colors are then retrieved by Fourier processing of the spectrally multiplexed hologram, as discussed in detail previously[48].

Figure 7c summarises a typical dataset obtained on a fixated Au-incubated HeLa cell where we highlighted a few "nanoparticle-like" and "cell-like" nano-objects. Comparing typical observations with the single-particle measurements (Figure 7a), suggests that the distinct AuNP RI-responses allow two-colour-based identification.

To gain systematic insight we first identify potential Au NPs over the entire 3D volumes by calculating so called Haar-features used in digital image processing[49]. The Haar-feature consists of subtracting the summed RI over a certain number of pixels (line, area or volume), from the sum over an adjacent region. This operation identifies the strong RI change around the position of Au NPs (Figure 7a). We perform this operation along the z-dimension. More specifically, at 505 nm we subtract the sum over 11 pixels from the adjacent 11 pixels, at 580 nm we introduce an additional 3 pixel gap between the regions. These values were selected empirically to yield the highest contrast. Thresholding at 0.008 (505 nm) and 0.01 (580 nm) RI contrast yields a list of possible particles for both wavelengths. On average, for the illumination wavelength of 580 nm around 250 particles are detected for an Au NP-incubated cell and 100 particles if no Au NPs are present. Using four samples each, we detect 966 particles (Au incubated) and 403 particles (not incubated). The number of initially detected particles at 505 nm is about twice as high, as the contrast of the particle response is weaker and thus more comparable with the background (data not shown).

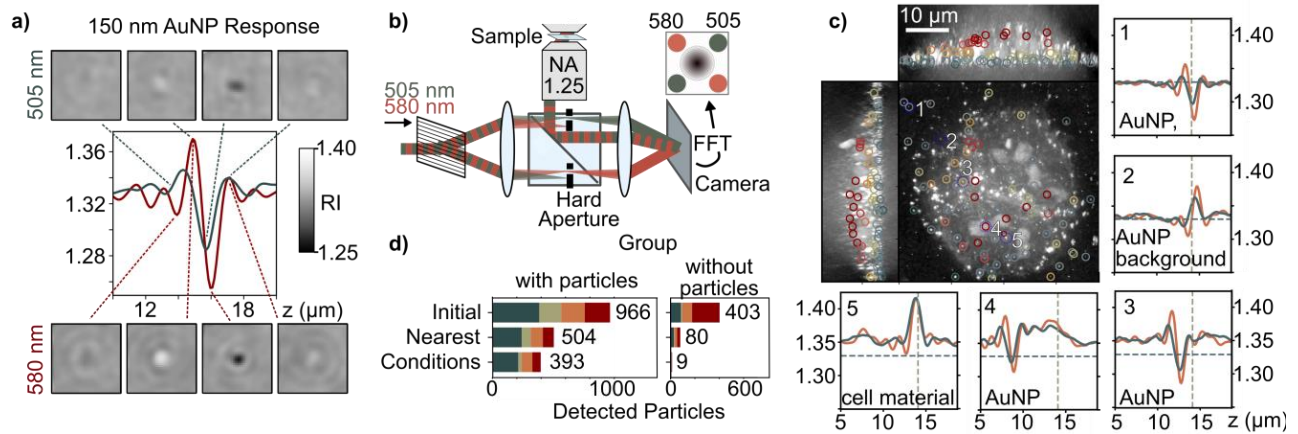

**Figure 7, Multiplexed two-colour tomography for robust 3D particle identification.** a) Rytov-extracted RI-signals for 150 nm AuNP at 505 nm (blue) and 580nm (red). b) Two reference-waves enable multiplexed hologram acquisition at 505 nm and 580 nm. c) Maximum RI projection values along x, y and z (greyscale) alongside detected NPs. The colour-scale indicates the z-position. Line plots show the RI along z for selected positions. The horizontal line indicates the RI of water and the vertical line the z-position of the cover glass. d) Statistics of four respective measurements with and without Au NPs using increasingly more rigorous filtering. Each colour represents one measurement.

To reduce the number of false positives we make use of the two-colour detection. First, we eliminate all particles that are not present in both detection channels and only keep those which are present in both, with a maximum lateral distance of 240 nm and axial distance of 660 nm. This step removes around 80 % of the false positives in the control group without Au NPs (Figure 7d, "Nearest"). To further reduce this number, we introduce additional conditions based on the typical RI responses (Figure 7a or Supplementary Information 8). These are, that the RI at 580 nm is smaller than at 505 nm and that the contrast of the Haar-feature at 580 nm is higher than at 505 nm. Further, the z-position of the minimum RI at 580 nm is smaller than at 505 nm, a feature also apparent in Figure 7a and possibly originating from the different focus of the objective lens [23]. Also, we set an absolute threshold to both features of 0.05 to remove clear outliers. Lastly, we remove particles that were present in the background image, by exploiting the fact that the Haar Features are reversed along z for these in comparison with the original particles. Using the two-colour approach, we obtain 2.25 particles per non-incubated sample, and 393 particles in the incubated sample. Reassuringly, a similar number of particles is eliminated from both samples, which suggests that our constraints selectively remove non-Au particles (Figure 7d).

## 4. Summary and conclusion

To summarise, we experimentally implemented high resolution, spectrally resolved, broadband ODT over bandwidths that are compatible with high temporal resolution experiments. We successfully recovered signals covering essentially the entire visible spectral range (500-650 nm), as validated by comparing to established, narrowband, spectral scanning alternatives. These capabilities directly enable highly temporally resolved experiments which have, thus far, been incompatible with ODT implementations. Further, we showed that multicolour tomography allows robust NP identification, based on known RI differences between biological and non-biological matter. Taken together, our results provide the necessary experimental framework for time-resolved 3D imaging of photoinduced RI changes in both synthetic as well as biological matter.

Moving forward, we identify two areas that require further improvement. Experimentally, the illuminating angles were created by a rotating grating. Besides being slow, this approach also allows only a cone geometry for illumination. By using diffractive optics, such as spatial light modulators or digital micromirror devices[25,50], the imaging speed could be greatly improved which would allow further improving the axial resolution[18]. Computationally, the Rytov approximation provided a good framework in this work (Supplementary Information 5). However, the implementations of algorithms (i) retrieving the real and imaginary part of the RI[19,21], (ii) mitigating artefacts of the missing cone problem to increase the fidelity of the 3D volume[17,41–43] and being applicable to scattering media[51] to resolve thick tissue is desirable. Given the current move towards biomedical imaging in the deep ultraviolet spectral range[52], we expect that algorithms for 3D RI-retrieval of highly scattering media as well as thick tissue will become available in the near future.

The methodology developed here bridges the gap between ODT and broadband ultrashort pulses and is directly applicable to pump-probe ODT with the only modifications necessary being to exchange the temporally chirped supercontinuum source employed here with a transform-limited temporally compressed source and to add an excitation pulse with controllable time delay. We expect that the ultrafast phototransient response will give a high contrast for nano particle detection [13] in otherwise highly scattering media and will permit to study complex time-resolved processes in three dimensions.

## 5. Acknowledgments

M.H. and G.C. acknowledge financial support from the Marie Skłodowska-Curie project 812992—"MUSIQ". F.V., M.L., G.C., and N.F.v.H. acknowledge funding from HORIZON- EIC-2021-PATHFINDEROPEN-01 project "TROPHY" (grant agreement no. 101047137). M.L. acknowledges support by the Spanish Ministry of Science, Innovation, and Universities (RTI2018-099957-J-I00) and was financially supported by the European Union (ERC, PIRO, Grant number: 101076859). Views and opinions expressed are however those of the author(s) only and do not necessarily reflect those of the European Union or the European Research Council. Neither the European Union nor the granting authority can be held responsible for them.

# Supplementary Information

1. Matching Spectral Phase of Reference and Signal Pulses

To achieve interference over the entire wavelength range of interest it is necessary to carefully balance the spectral chirp of the "Reference" and "Signal" arms of the broadband optical diffraction tomography (ODT) setup (Figure 2). We mimic the highly dispersive elements present in the microscope by adding dispersive components to the reference path, mainly in the form of achromatic lenses, to approximate the highly dispersive materials used in high numerical aperture microscope objectives. A pair of wedge prisms fine-adjusts the chirp.

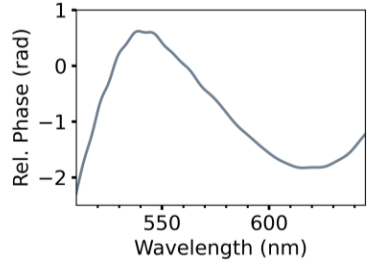

**Figure SI 1:** Relative phase (in radians) after balancing the dispersion in the reference arm of the ODT setup.

To quantify the chirp difference between the two interferometer arms, we rely on spectral interferometry by placing a commercial spectrometer (*Ocean Optics*) at the position of the camera. As such, we directly record spectrally resolved interference between reference and signal waves. Fourier analysis[1,2] directly retrieves the difference between the spectral phases of the pulses, which corresponds to a difference between the arrival times of their different frequency components. We add dispersive elements (lenses, windows) to the reference path until the relative spectral phase becomes as flat as possible.

Figure SI 1 shows the obtained wavelength dependent relative spectral phase where the residual higher order dispersion is, most likely, due to material differences where especially high numerical aperture objectives often contain highly dispersive, specialized, optical glass.

2. Reconstruction Procedure for Rytov and Born Approximation

The following reconstruction is carried out for each wavelength separately. We start with the complex extracted fields $f_0(x, y, \vartheta)$ and $f_{bg}(x, y, \vartheta)$. For the Rytov approximation, we normalize the as-retrieved fields [3]:

$$f_{norm,\vartheta}(x,y) = \frac{f_0(x,y,\vartheta)}{f_{bg}(x,y,\vartheta)} = A_\vartheta(x,y) \exp[i\, \varphi_\vartheta(x,y)],$$

and then apply the complex logarithm:

$$f_{Ryatov,\vartheta}(x,y) = \log\left(f_{norm,\vartheta}(x,y)\right) = i\, \varphi_\vartheta(x,y) + \log(A_\vartheta(x,y)).$$

For the Born approximation we use[3]:

$$f_{norm,\vartheta}(x,y) = \frac{f_0(x,y,\vartheta) - f_{bg}(x,y,\vartheta)}{f_{bg}(x,y,\vartheta)}.$$

We omitted the illumination angle, $\vartheta$, to simplify the discussion but note that the operation has to be performed for all fields. Then the process is for both approximations the same. A 2D FFT yields the k-space distribution,

$\bar{f}_{Ryatov}(k_x, k_y)$, with the illumination $k$ being shifted to the center of the k-space, or DC, due to the background normalization (inset Figure SI 2a). The next step is to use knowledge of the spatial frequencies $k_x$, $k_y$ as well as the illumination spatial frequencies $k_{x0}$, $k_{y0}$ to reconstruct the Ewald sphere. We require knowledge of $k_{Ewald}(k_x, k_y)$, the $k_z$-component on the Ewald sphere, which is accessible via the refractive index of the surrounding medium, $n_m$, and the wavelength $\lambda$, $k_0 = \frac{2\pi}{\lambda}$:

$$k_{Ewald}(k_x, k_y) = \sqrt{(n_m k_0)^2 - (k_x + k_{x0})^2 - (k_y + k_{y0})^2}$$

if $(k_x + k_{x0})^2 + (k_y + k_{y0})^2 < (k_0 * n_m)^2$, otherwise zero,

where we re-introduced the original illumination information $k_{x0}$ and $k_{y0}$ to move the Ewald sphere on top of the circle in k-space (inset Figure SI 2a). Once $k_{Ewald}(k_x, k_y)$ is determined, we scale the Rytov-retrieved k-space distributions as:

$$\bar{f}_{Ewald}(k_x, k_y) = \frac{i}{\pi} k_{Ewald}(k_x, k_y) \, \bar{f}_{Ryatov}(k_x, k_y).$$

Finally, each of the values is assigned to its position within a 3D scattering potential $\bar{F}(k_x, k_y, k_z)$ of the object of interest, which lies on the shifted semi-sphere: the Ewald sphere. The $k_z$ position is inferred as follows:

$$k_z(k_x, k_y) = k_{Ewald}(k_x, k_y) - k_{z0}.$$

This process is repeated for all angles $\vartheta$ to fill the 3D potential, as shown in Figure SI 2b. For points which are covered by multiple acquisitions, using different acquisition angles, mean values are calculated. Finally, an inverse 3D FFT retrieves the distribution in real space, which is converted from scattering potential to the refractive index (Figure SI 2c):

$$RI(x, y, z) = n_m \sqrt{1 - F(x, y, z) * \left(\frac{\lambda}{2 * \pi n_m}\right)^2}.$$

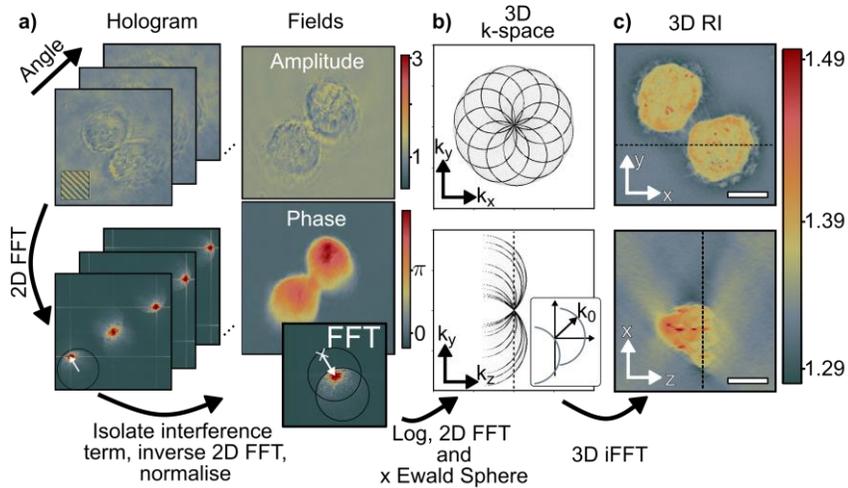

**Figure SI 2, From raw holograms to 3D tomograms.** a) A set of holograms (with and without sample) is recorded at different illumination angles and then Fourier-processed to retrieve normalized amplitude and phase images. The top inset shows the interference fringes of the hologram. The bottom inset corresponds to the 2D FFT of the normalized image. The white arrows show the (lateral) illumination beam in k-space and the shift due to normalization. b) The object's scattering potential in 3D k-space is sampled using the 2D FFT of the normalized fields and the corresponding Ewald sphere of each illumination angle.

Only a few subsets are highlighted for clarity. The inset shows the shift of the Ewald sphere for an illumination with k-vector $\underline{k}_0$. c) An inverse 3D FFT of the Ewald sphere retrieves the 3D, real space, RI of the object. Dotted lines represent the respective image slices. Scale bar: 10 μm.

## 3. Extracting Spectra from Pulse-Pair recorded Holograms

As described in the main text in Section 2.2 we record holograms for which the spectrum is modulated by a cosine square to enable Fourier transform based spectrally resolved imaging. For each modulation frequency $\nu$ we record one hologram.

We then retrieve the spectrum on a pixel-by-pixel basis. We define H(x, y, $n_\vartheta$) as the value of the pixel at position x and y and $n_\vartheta$ as the n-th acquisition for the modulation $\vartheta$. Then we retrieve one-dimensional arrays over all $n_\vartheta$ acquisitions which we call $f_{xy}$.

Our original acquisition has 24 modulation frequencies, covering a 0- 40 fs temporal delay range. Before performing the fast Fourier transformation (FFT) over $f_{xy}$, we invert $f_{xy}$, delete it's first component and append it to the original $f_{xy}$, as shown in Figure SI 3a. Following this operation, we perform the one-dimensional FFT over the appended array of $f_{xy}$, yielding the spectral information of interest (Figure SI 3b). The wavelength region of interest is highlighted.

Keeping in mind that we use a carrier frequency with wavelength $\lambda_{carrier}$, we retrieve its frequency as $\nu_{carrier} = \frac{c}{\lambda_{carrier}}$. We calculate the spectral modulation corresponding to a time delay, τ, as $\vartheta$: $M(\nu, \tau) = |\cos((\nu - \nu_0)\pi\tau)|^2$.

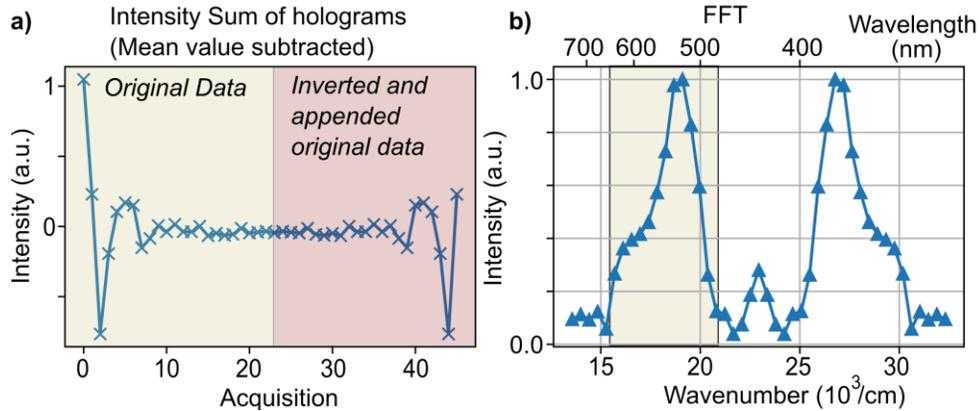

**Figure SI 3: Pixelwise FFT extraction for pulse pair spectroscopy.** a) Prior to the fast Fourier transformation (FFT) the one-dimensional array (orange) is inverted and appended to the back of the original array (blue). As an example we plotted here the absolute value, whereas the fields are complex. b) Resulting FFT, the spectral region of interest is highlighted in orange.

## 4. Retrieving k₀ₓ, k₀ᵧ and k₀ᵤ

Tomographic reconstruction requires precise knowledge of the illumination angle, or k-vectors. We use the background fields, e.g. images recorded in an empty region of the sample, to retrieve the original illumination vectors. We measure the phase of each illumination angle (two-dimensional array) and unwrap it to retrieve a phase which allows calculating the k-vectors as $k_{x0/y0} = \frac{\Delta\varphi_{x/y}}{p}$, with $\Delta\varphi_x$ and $\Delta\varphi_y$ being the phase change per pixel and $p$ the physical pixel size at the sample plane. The $k_z$-component is retrieved via $k_{z0} = \sqrt{(k_0 * n_m)^2 - k_{x0}^2 - k_{y0}^2}$, with $k_0 = \frac{2\pi}{\lambda}$, $\lambda$ the wavelength of the incoming light and $n_m$ the refractive index of the surrounding medium. Figure SI 4 shows the retrieved k-vectors and angles of a pulse-pair scan for all wavelengths.

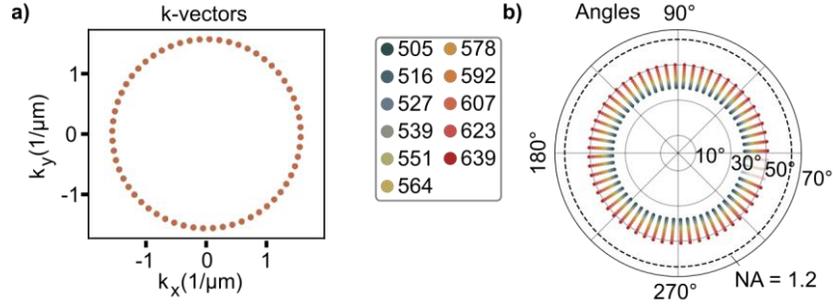

**Figure SI 4: k-vector retrieval.** a) k-vectors retrieved for all wavelengths used in FFT approach (505 nm to 639 nm). b) Illumination angles extracted for all wavelengths.

## 5. Missing Cone Problem in ODT

The limited illumination angles allow to fill the k-space of the object only partially, as apparent in Figure 4b. This is known as missing cone problem. The consequences can be seen in the reconstruction in Panel 4c as a slight blurring along the z-coordinate and RI values below water at the edges of the object. Furthermore, the overall RI value is slightly underestimated. To this end, many iterative solvers exist trying to circumvent this problem[3–5], or to make ODT applicable outside the Rytov approximation for scattering media[6]. These algorithms mitigate the missing cone problem and increase the fidelity of the retrieved 3D objects. However, they do not gain further real information, e.g. record information beyond the diffraction limit. Thus, in the context of this article, which mainly focuses on implementing ODT with broadband pulses, we stick to the Rytov approximation as it (i) best visualizes the recorded experimental data (only a two-dimensional phase unwrapping algorithm is necessary) without the use of complex iterative solvers, which could introduce their own artefacts, and (ii) is sufficiently accurate for the data presented here (ignoring the artefacts due to the missing cone problem).

## 6. Illumination via Wavelength-Scanning

In the context of the proof-of-principle experiments presented in Figure 3 of the main text we implement a wavelength scanning approach. The spatial light modulator sweeps illuminations with 10 nm bandwidth. Figure SI 5 shows the spectra after the Fourier filter recorded with a commercial spectrometer (*Ocean Optics*). In order to remove the leakage of the SLM, one acquisition with all wavelengths blocked is acquired and in a post-processing step subtracted from the other acquisitions.

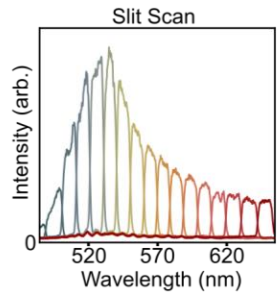

**Figure SI 5: Recorded spectra used for wavelength-scanning.**

## 7. Wavelength Dependent k-Space and Resolution

ODT allows extending the spatial resolution. Following reference[7], the resolution is extended up-to two fold $k_x = k_0 * n_m * \sin(\theta)$, depending on the illumination angle $\theta$ between axial and lateral illumination k-vectors. Thus, the resulting lateral resolution is:

$$d_{lat} = \frac{1}{2\,k_{x,NA} + 2\,k_x}$$

with $k_{x,NA} = k_0\,n_m \sin(\theta_{NA})$, $\theta_{NA}$ is maximum angle defined by NA ($NA = \sin(\theta_{NA})\,n_m$).

If the illumination is equivalent to $\theta_{NA}$, then $d_{lat} = \frac{\lambda}{4NA}$ which is the maximum resolution limit. Analogously, for an illumination with cone-geometry, the axial resolution reads as

$$d_{ax} = \frac{1}{n_m\,k_0\,(1 - \cos(\theta_{NA}))}.$$

Figure SI 5 shows the wavelength dependent, calculated, spatial resolutions.

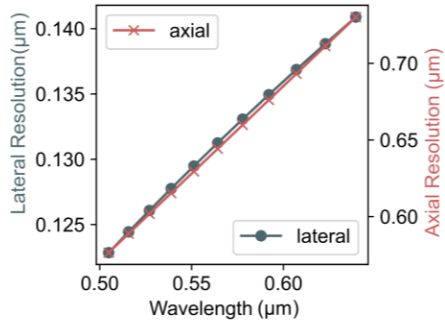

**Figure SI 5: Theoretical lateral and axial resolution in ODT with cone-illumination calculated for the experimental parameters and illumination angles from Figure SI 4.**

## 8. Spectral Response of 150 nm Au Nanoparticles

To further validate the setup, we recorded broadband tomographic images of gold nano particles (Au NPs) with a diameter of 150 nm immobilized on glass. In contrast to the off-resonant HeLa cells, Au NPs show a strongly wavelength dependent RI in the visible spectral range: the RI of bulk gold strongly decreases with wavelength[8]. Figure SI 6a shows an xy-plane cut in the vicinity of a cover glass surface of a reconstructed Au NP tomogram obtained with a pulse pair. We observe multiple Au NPs, immersed in water and immobilized onto a glass coverslip, that show a RI that is lower than water, as expected based on the RI of bulk gold. As previously, we examine the wavelength dependence of the RI (Figure SI 6b) where we note a marked decrease with increasing wavelength when analyzed in the respective tomogram plane. However, the xz-projections through the center of the representative Au NP show a RI increase followed by a decrease, especially at longer wavelengths. To rationalize this observation, we concentrate on averaged RI line cuts, obtained by combining the data of all Au NPs (Figure SI 6c). The RI at 607 nm shows a derivative profile with a decrease to a minimum of around 1.26 as well as an increase to up to 1.39. Based on the bulk RI of gold, one might have expected RI values of around 0.5-0.8. Given that the particles employed are strongly scattering, reconstruction-artefacts due to loss of information are a likely cause for this discrepancy. A further reason for the only partial recovery of the RI values is the small size of the Au NPs compared to the 130 / 650-nm lateral/axial point spread function of the system, which essentially averages the RI response of the NPs with that of the surrounding medium.

To ensure that no bandwidth-associated artefacts are present, we first compare the particles' responses as measured via a slit scan with the pulse-pair methods. Figure 6 SIc,d highlight that both methods yield near-identical results. In other words, the RI-oscillations are not due to an experimental error. To gauge the impact of the reconstruction procedure, we compare the RI values as obtained via the Rytov approximation (Figure SI 6d) to the one obtained following a prominent alternative: the Born approximation[3] (Figure 6 SI e). The latter is considered valid for sample-induced phase changes below $\frac{\pi}{2}$ [3], a condition that is fulfilled by the Au NPs. Indeed, the positive contribution vanishes with the Born approximation, which suggests that it is a more accurate reconstruction procedure for the Au NPs. From a computational perspective, more advanced algorithms are likely to yield even better results [4,5,9,10]. Another reason for spectral differences could be the k-space filling, which is wavelength dependent.

To validate the effect of the wavelength dependent k-space filling on our data, we perform a simple simulation. First, we save for each wavelength a binary mask in the k-space of points which our imaging system retrieves (Figure SI 6f). Then, in real space, we create a 3D array with the RI values of water and set a single voxel in the center to a RI value of 0.7, similar to Au. The pixel size in the experiment was 122 nm, which is close to 150 nm Au diameter. Therefore, we use the same pixel size, and perform analogously to the experiment a 3D FFT, remove the frequencies we would also loose in the experiment (missing cone), and perform the inverse FFT. The results are shown in Figure SI 6g. In contrast to the real data, the spectral missing cone problem causes a increase towards higher wavelengths. Further, it has a symmetric point spread function. Thus, we can conclude that (i) the decrease in the RI with higher wavelengths is due to the material and (ii) the artefact of a larger RI than water with higher wavelengths is due to the Rytov approximation.

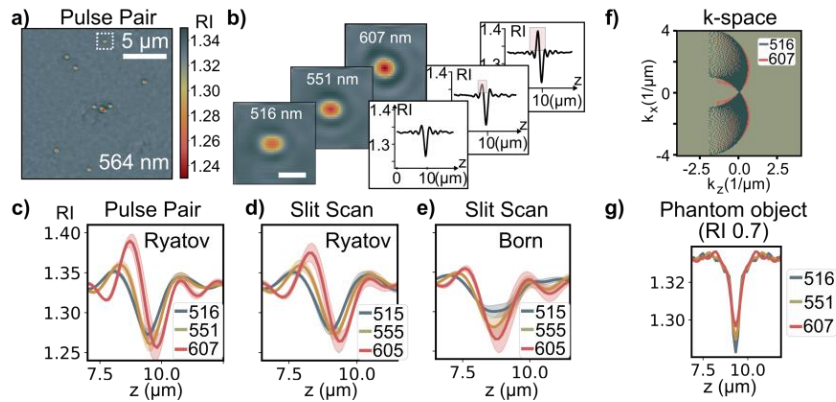

**Figure SI 6, Wavelength resolved 3D-reconstructions of 150 nm Au NPs using different approaches. a) Image of Au NPs on top of a glass substrate and immersed in water obtained using the pulse pair method, the Rytov approximation and a wavelength of 564 nm. b) Image of one particle at different wavelengths together with a line cut of the RI along z. Scale bar: 500 nm. c) Averaged z-dependence of RI for the particles in a) acquired with the pulse pair method. Shaded areas correspond to two standard deviations. d) Averaged RI from NPs acquired with the slit scan method. e) Same NPs as in d) but reconstructed with Born approximation.**